\documentclass[twoside]{article}
\usepackage{fleqn,espcrc2,amsfonts,epsfig}

\newcommand{\id}{{\rm 1\kern-.12em
\rule{0.3pt}{1.5ex}\raisebox{0.0ex}{\rule{0.1em}{0.3pt}}}\,}

\def \be {\begin{equation}}
\def \ee {\end{equation}}                                         
\def \bea {\begin{eqnarray}}
\def \eea {\end{eqnarray}}                                         
\def \MT {\mathcal T}

\newcommand{\ba}{\begin{array}}
\newcommand{\ea}{\end{array}}

\newcommand{\pref}[1]{(\ref{#1})}

\begin{document}


\hyphenation{author another created financial paper re-commend-ed Post-Script}


\title{
{\vspace{-6em} \normalsize                                             
\hfill \parbox{23mm}{\sc DESY 99-137}}\\[25mm]
\vspace{-1cm}
Global anomalies in Chiral Lattice Gauge Theory\thanks{based on the
talks given by O. B\"ar and I. Campos. Presented at the
International Symposium on Lattice Field Theory, June 29 - July 3, 1999,
Pisa, Italy.}}

\author{Oliver B\"ar and Isabel Campos \\
\vspace{0.4cm} 
{{\sl Deutsches Elektronen-Synchrotron},
 {\rm Notkestrasse 85, 22603 Hamburg (Germany)}} }

\begin{abstract}

As first realized by Witten an $SU(2)$ gauge theory coupled to a single
Weyl fermion suffers from a global anomaly. 
This problem is addressed here in the context of the recent developments
on chiral gauge theories on the lattice. We find Witten's
anomaly manifests in the impossibility of defining globally a fermion 
measure that reproduces the proper continuum limit.
Moreover, following Witten's original argument, we check numerically
the crossing of the lowest eigenvalues of Neuberger's operator along
a path connecting two gauge fields that differ by a topologically
non-trivial gauge transformation.

\vspace{1pc}
\end{abstract}

\maketitle

\section*{Introduction}

In \cite{WITTEN} Witten pointed out the mathematical inconsistency of an
$SU(2)$ gauge theory coupled to only an odd number of Weyl fermions.

The inconsistency arises when one consi\-ders the behaviour of the effective
action under topologically non-trivial gauge transformations
($\pi_4[SU(2)] = {\mathbb {Z}}_2$), i.e. those that cannot be continuously
deformed to the identity mapping.

Let $g$ be one of those non-trivial mappings.
The gauge fields $A_{\mu}$ and its gauge transformed 
\begin{equation}
A^{\rm g}_{\mu} = {\rm g} A_{\mu} {\rm g}^{-1} + {\rm g} \partial_{\mu} 
{\rm g}^{-1}  \ ,
\end{equation}
are not connected by some smooth gauge transformations. 
However, they are connected
in the space of all gauge fields since it is a vector space. Thus
\begin{eqnarray}
A^t_{\mu} = (1-t)A_{\mu} + tA^{\rm g}_{\mu}, \quad t \in [0,1]   \ ,
\label{PATH}
\end{eqnarray}
is a well defined potential which interpolates between $A_{\mu}$ and
$A^{\rm g}_{\mu}$. 
Witten showed that along this trajectory an odd number
of eigenvalues of the square root of the Dirac operator crosses zero, leading to
a switch of sign in the fermionic determinant \cite{WITTEN}.
The theory is thus ill-defined because the fermion determinant cannot be 
defined in a gauge invariant and smooth way.

Afterwards, the Witten anomaly has been established by other arguments.
In particular, the change in the phase of the effective action under $SU(2)$
gauge transformations can be calculated using the so-called
{\sl embedding technique} \cite{ELITZUR}. Here one uses the fact
that $SU(2)$ can be embedded in another group with a trivial $\pi_4$,
say $SU(3)$. The gauge fields $A_{\mu}$ and $A^g_{\mu}$ are then connected
by some smooth family $\Omega$ of
gauge transformations in $SU(3)$. In this context the 
change in the phase of the effective action is given by \cite{WITTEN2}
\begin{equation}
\Delta \Gamma_{\rm eff}(\Omega d \Omega) =
i\frac{-i}{240\pi^2} \int {\rm tr} [(\Omega d \Omega)^5] = i\pi \ ,
\label{CS}
\end{equation}
leading to the switch of sign in the fermionic determinant.

During the last years encouraging progress has been made to formulate
chiral gauge theories on the lattice as has been summarised in
\cite{PLENARY}. Any proper lattice formulation of an $SU(2)$
theory coupled to one doublet of chiral fermions should exhibit
the Witten anomaly. This has been stu\-died in the context of the
overlap formalism \cite{NEU2}. Our purpose is showing how global anomalies
arise in the recently developed action formalism \cite{ABEL,SU2}. 

\section*{Weyl fermions on the lattice}

Exact chiral symmetry \cite{SYMMETRY} can be achieved on the lattice 
provided the Wilson-Dirac operator $D$ fulfils the Ginsparg-Wilson relation 
\cite{GW}.

Having exact chiral symmetry allows us to split the fermion
fields in left- and right-handed, independently transforming 
components \cite{SYMMETRY,NIED,NARA}. In particular, 
we can restrict ourselves to the left-handed fields imposing the constraints
\bea
\label{CONS1}
\hat{P}_- \psi &=& \psi  \ ,\\
\bar{\psi}P_+ &=& \bar{\psi} \ .
\eea
One of the new features of the present approach 
is that $\hat{P}_- = \frac{1}{2} (1 - \gamma_5(1 -aD))$ depends on 
the gauge field. Let us discuss some of its geometrical implications.

Consider a path in configuration space
$U_{\rm t} (x,\mu)$, where $t \in [0,1]$ is the path parameter.
We define an unitary operator $Q_{\rm t}$ through the differential equation
\bea
\partial_t Q_t = [\partial_t P_t,P_t] Q_t \ , \qquad\mbox{$Q_0 = \id$} \ ,
\label{DE}
\eea
where $P_{\rm t} \equiv {\hat P} |_{U=U_{\rm t}}$.
The operator $Q_{\rm t}$ is such that 
$P_{\rm t} Q_{\rm t} = Q_{\rm t} P_0$. In this way $Q_{\rm t}$ is the
transporter of $P_{\rm t}$ along the path.

One key point here is that if the path is a closed loop,
the operator $Q_1$ is not necessarily the identity map. Indeed, 
$Q_1 \neq \id$ is an indication of a non-trivial bundle structure of
the gauge field. As a measure for this let us define the quantity
\be
\MT = \det [1 -P_0 + P_0 Q_1 ] \ ,
\ee
for all closed loops. 
Using the unitarity of $Q_{\rm t}$
one easily proves that $\MT$ is a phase. 
In particular, for gauge fields in $SU(2)$, using charge conjugation 
symmetry and the reality properties of the $SU(2)$ representations 
we have $\MT = \pm 1$.

Finally we remark that the composition law 
\be
\MT_{[\Gamma_1 \circ \Gamma2]} = \MT_{\Gamma1} \, \MT_{\Gamma_2} 
\label{COMPO}
\ee
holds for two closed loops that have the same starting point.

\section*{Fermionic Measures}

The action of the classical gauge theory coupled 
to a single Weyl fermion reads
\begin{equation}
S_{F,L} = a^4 \sum_x \bar{\psi}(x) [P_+ D \hat{P}_- \psi](x) \ .
\end{equation}
In order to set up the quantum theory we have to define a measure
for the fields in the path integral. Here the difficulty arises 
due to the gauge field dependence of the constraint (\ref{CONS1}).
In fact, an infinitesimal deformation $\eta_{\mu}(x)$ 
of the gauge field $U(x,\mu)$, induces a change in the phase
of the fermionic measure given by the so-called measure term \cite{SU2}
\bea
{\mathfrak L}_{\eta} = i \sum_j (v_{\rm j}, \delta_{\eta} v_{\rm j})
\equiv a^4 \sum_x \eta^c_{\mu}(x)j^c_{\mu}(x)  \ ,   
\label{MEASU}
\eea
where $c$ is the colour index and
$\{v_{\rm j}\}$ is a basis of left-handed fields at $U(x,\mu)$.

We are then left with a gauge dependent
phase ambiguity which has to be fixed in order to achieve
the gauge invariance and the locality of the effective action.
In \cite{ABEL,SU2} the problem is solved  
by choosing a current $j_{\mu}(x)$, local and gauge invariant 
function of the gauge fields, and such that an integrability
condition is fulfilled.
This condition is formulated in terms of the Wilson line
\be
{W} = \exp \left\{ i \int_0^1 dt {\mathfrak L}_{\eta} \right\} \ ,
\ee
where $a\eta_{\mu}(x) \;=\;\partial_tU(x,\mu)U(x,\mu)^{-1}$. $W$ measures
the total change of phase along a given path $U_{\rm t} (x,\mu)$ in the set 
of gauge fields.

The integrability condition states that for all closed loops in
configuration space the Wilson line must satisfy
\be
W=\MT \ .
\label{IC}
\ee
Let us discuss the meaning of this condition.
A local and gauge invariant current defines $W$ only locally. 
In (\ref{IC}) it is pointed out that the 
current must take into account the global geometry of the bundle
underlying the gauge field.
Global anomalies arise when (\ref{IC}) is not satisfied. 

In the classical continuum limit, requiring a local and gauge
invariant effective action implies \cite{SU2}
\be
j_{\mu}(x) = 0  + {\mathcal O}(a) \ .
\label{JCC}
\ee
Therefore $W=1  + {\mathcal O}(a)$ in this limit and the integrability
condition can only be satisfied if $\MT~=~1$~for all closed loops.

In the next section we will show that in the $SU(2)$ theory 
with a single Weyl fermion there are closed loops in configuration 
space on which $\MT = -1$. Therefore the anomaly arises because the proper
classical continuum limit cannot be reproduced.

\section*{SU(2) Global Anomaly}

Let $g$ be a non-trivial $SU(2)$ gauge transformation,
and consider its lattice version acting on the classical vacuum, $U(x,\mu)=\id$.

There are three different paths in configuration space connecting
the classical vacuum with its gauge transformed, $g(x) g(x+a\mu)^{-1}$:
\begin{itemize}
\item{$\Gamma_1 \equiv$} $[g(x)^tg(x+a\mu)^{-t}]$ pure gauge $SU(2)$
\item{$\Gamma_2 \equiv$} $[\Omega(t,x) \Omega(t,x + a \mu)^{-1}]$  
   pure gauge $SU(3)$
\item{$\Gamma_3 \equiv$} $(g(x) g(x+a\mu)^{-1})^t$
\end{itemize}
$\Gamma_1$ has no continuum limit. $\Gamma_2$ is the lattice analogue
of the $SU(3)$ pure gauge path defined in the introduction and
$\Gamma_3$ is a lattice version of Witten's path (\ref{PATH}).

Our aim is computing $\MT$ on the
closed loop $[\Gamma_3 \circ - \Gamma_1]$.
Using the composition law (\ref{COMPO}) we split $\MT$ in the following way
\begin{equation}
\MT_{[\Gamma_3 \circ - \Gamma_1]} = \MT_{[\Gamma_3 \circ - \Gamma_2]} \,
\MT_{[\Gamma_2 \circ - \Gamma_1]} \ .
\end{equation}
Along pure gauge loops we have, 
\bea
 {\mathcal T} = \exp \left\{-i\int_0^1 \: dt \sum_x \omega_t^a(x)
{\cal A}^a(x)   \right\}  \ ,
\eea
where ${\cal A}^a(x)$is the anomaly on the lattice.

For the vacuum configuration the anomaly vanishes because
it is invariant under translations but
also odd under parity, ${\cal A}^{a}(-x) = - {\cal A}^a(x)$. 
That means $\MT_{[\Gamma_2 \circ - \Gamma_1]} = 1$.
This argument holds for the vacuum
in $SU(N)$. However, in $SU(2)$ the anomaly is identically 
zero in any configuration.

We are then left with the calculation of $\MT$ for the
loop $[\Gamma_3 \circ - \Gamma_2]$. To deal with it we
convert the line integral into a surface integral. On this
surface the vector potential depends on two parameters ($t,s$).
It can be shown \cite{FUTURE} that
\be
\partial_{\rm s} \ln \: {\mathcal T} = 
\int_0^1 dt \: {\rm Tr} {\hat P}_- \, [\partial_s {\hat P}_-, \partial_t {\hat P}_-] \ .
\label{EQCON}
\ee
The r.h.s. of equation (\ref{EQCON}) can be expanded in powers of the
lattice spacing $a$ and to leading order one finds
\be 
\partial_{\rm s}\ln{\mathcal T} \\
=  -i c_2 \int_0^1 dt \int d^4 x \, d^{abc}_R \epsilon_{\mu \nu \rho \sigma
} 
\eta^a_{\mu} \xi^b_{\nu} F^c_{\rho \sigma}
\label{ASYM}
\ee
Here we take over the notation of \cite{SU2}. The deformations
$\eta$ and $\xi$ corresponds to the $t$ and $s$ directions respectively.
In (\ref{ASYM}) we substitute our parameterisation of the vector
potential on the surface.
Integrating over $s$ we finally end up with the integral 
(\ref{CS}) and find $\MT_{[\Gamma_3 \circ - \Gamma_2]} = -1$.
Altogether we get $\MT_{[\Gamma_3 \circ - \Gamma_1]} = -1$.

We want to point out that our proof is not restricted to have
the classical vacuum as starting configuration 
(although it would be sufficient for the theory to be inconsistent).
Since $\MT$ is a homotopy invariant, smooth deformations of the loop
cannot change its value.


\section*{Witten Anomaly}

Next we want to show that the global anomaly we discovered 
in the previous section
can be brought into a form that its equivalence to Witten's anomaly
\cite{WITTEN} is transparent. It will turn out that if we restrict
ourselves to a  
real fermion determinant, $\MT=-1$ implies a change of sign along the
loop $[\Gamma_3 \circ - \Gamma_1]$.

To start with consider the function
\bea
f(t)  =  \det \, (1-P_+ + P_+D_tQ_tD_0^{\dagger}),
\eea
that is a smooth function of $t$ for smooth paths. In addition it is
real if the gauge group is $SU(2)$. For closed loops that function
satisfies
\bea
f(0)\,>\,0\,, & & f(1) = {\cal T}f(0) \ ,
\eea
if $D$ has no zero mode at the starting point. This implies  that $f$
passes through zero an odd number of times for $0\leq t\leq 1$ if and
only if ${\cal T}=-1$.

Next one can show
\bea\label{f_squared}
f^2 (t) =  \det D_t \det D_0^{\dagger}.
\eea
If $D^{\dagger}=\gamma_5 D \gamma_5$, the eigenvalues of $D$
come in complex conjugate pairs, i.e.
\bea
\det D=\prod\lambda_i\lambda_i^* \ .
\eea
According to \pref{f_squared} a
passing through zero of $f(t)$ at $t_0$ implies a passing through zero
of an odd number of
eigenvalues $\lambda_i(t)$. One can prove this by expanding
both $f$ and $\lambda_i$ around $t_0$.
Hence we have found that for any given
loop in field space, starting at a point where $D$ has no zero modes,
an odd number of pairs of eigenvalues of $D$ cross zero if and only if
$\MT = -1$.

Hence we find the same behaviour of the eigenvalues that Witten proved in
\cite{WITTEN} using
the Atiyah--Singer index theorem and that he used to conclude a change
of sign of the fermion determinant.
We can find the same here. In general, the fermion determinant is
given by \cite{SU2}
\be\label{formula_det}
\det M_t\det M_0^{\dagger} = f(t) \, W^{-1} \ .
\ee
If we define $j_{\mu}\equiv1$ we end up with a real fermion
determinant all along the path. According to the properties of $f$, we find that
$\det M_1 \det M_0^{\dagger}$ is negative. Therefore $\det M_t$ changes sign
along a loop with $\MT=-1$. 

Note however, that a change of sign is closely connected to the
definition of a real fermion determinant. If we allow for
complex values too and define a
current $j_{\mu}$ such that $W$ equals $-1$ at the end of the
loop, the fermion determinant is single valued. In
that case the anomaly shows up in the disguise of a conflict with the
correct continuum limit of $j_{\mu}$ (\ref{JCC}) and $W$.

\section*{Spectral flow of Neuberger's operator}

In the last section we have seen that ${\cal T} =-1$ implies that an
odd number of pairs of eigenvalues of $D$ cross zero. On a finite
lattice it is possible to confirm this crossing by a numerical
computation of the spectral flow of an appropriate lattice Dirac
operator, as we will see now.

Neuberger's operator \cite{Neuberger98a} is an example for a Dirac
operator that satisfies the Ginsparg-Wilson relation. 
It is explicitly given by 
\bea\label{def_Neub_op}
aD & = & 1-A(A^{\dagger}A)^{-1/2}\,,\\
 A & = & 1 - aD_w\,.
\eea
$D_w$ denotes the usual Wilson--Dirac operator. 
The eigenvalues $\lambda_j$ of $aD$ lie on a unit circle around 1 in
the complex  plane. 
They can be parameterised by an angle $\theta$ according to
\bea\label{def_ev}
\lambda_j & = & 1 -e^{i\theta_j}\,.
\eea
Furthermore the eigenvalues come in complex conjugate pairs
$\lambda_j,\lambda_j^*$ due to $D^{\dagger}=\gamma_5 D \gamma_5$. 

We are interested in the  spectral flow of Neuberger's operator along
the loop $\Gamma = [\Gamma_3\circ-\Gamma_1]$. According to the last
section we expect an odd number of eigenvalues that cross zero and go
over into its complex conjugate value.

Along $\Gamma_1$ the eigenvalues are constant as a function of $t$
because it is a path of gauge transformations. 
The non--trivial part is along $\Gamma_3$ where we numerically
computed the eigenvalues.
This has been done in two steps. 
First of all we computed the low lying eigenvalues of the hermitian operator
$aD^{\dagger}aD$.

They are given by the squared magnitude $|\lambda_j|^2$ of the
eigenvalues \pref{def_ev}.  
In a second step we calculated the imaginary part of $\lambda_0$,
i.e. the eigenvalue with the smallest imaginary part at the beginning
of the path. 
This is sufficient for our purpose. 
Before discussing the results let us make 
some remarks concerning the numerical computation itself.

We used a power series expansion into Chebyshev polynomials for the
inverse square root of $A^{\dagger}A$ in \pref{def_Neub_op}. 
The Conjugate Gradient algorithm \cite{Bunk94a,Kalkreuther96a} has been
employed for the computation of 
the eigenvalues. In that way both the truncation
and the numerical error are theoretically well under control.

For numerical reasons the path we used differs slightly from
$\Gamma_3$. 
We did not start at the classical vacuum configuration but at a
constant gauge field instead. 
This results in a gap of the spectrum at $t=0$ and $t=1$ and is
advantageous in the numerical computation. 

\begin{figure}[t]
\epsfysize=7.5cm
\epsffile{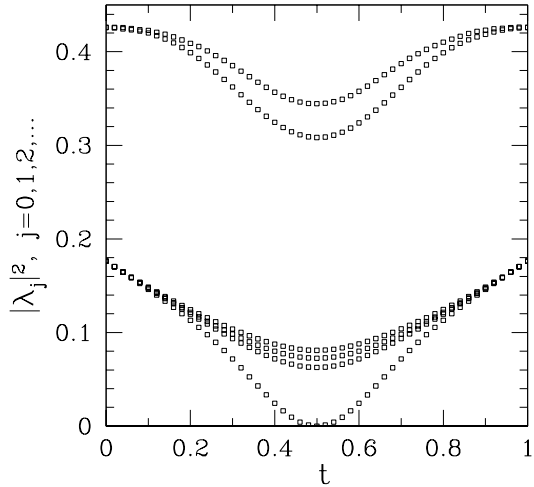}
\caption{\small The lowest six eigenvalues of
  $aD^{\dagger}aD$ on a $8^4$ lattice. The total error is smaller than the
  size of the data points.} 
\label{fig1}
\end{figure}

\begin{figure}[t]
\epsfysize=7.5cm
\epsffile{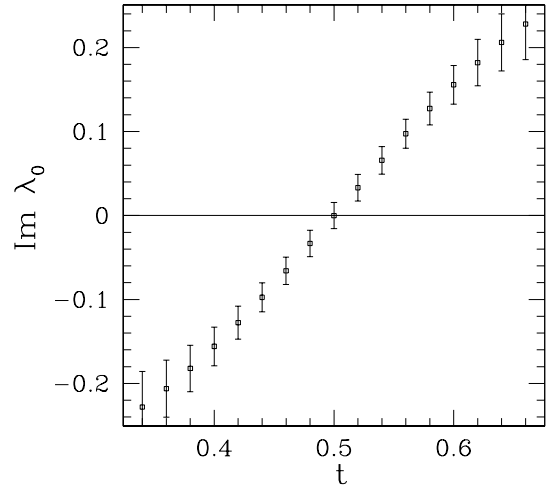}
\caption{\small The imaginary part of $\lambda_0$. The error bars
  incorporate both truncation and numerical error.} 
\label{fig2}
\end{figure}

Figure \ref{fig1} shows the first six lowest lying eigenvalues of
$aD^{\dagger}aD$ as a function of the path parameter $t$.
The obvious symmetry of the spectrum is due to our path
parameterisation and the symmetry properties
of the particular $g(x)$ we used. 
The magnitude of all but one eigenvalue is unequal to zero for the whole
path. 
Only the lowest eigenvalue becomes zero for $t=0.5$.
Hence only the imaginary part of $\lambda_0$ may cross zero and changes sign.  
To see if it really changes sign we calculated directly the imaginary
part of $\lambda_0$.
The result is shown in
fig.~\ref{fig2} and indeed we find a crossing. 
Hence only one pair of eigenvalues crosses zero and we numerically
confirmed a spectral flow that we have expected.

The change of sign of the fermion determinant can now be shown very
explicitly. In terms of the angles $\theta_j$ the fermion determinant
reads as
\bea\label{def_detM}
\det M_t & = & \prod_j\,2\,\sin\frac{\theta_j(t)}{2}.
\eea
Here we fixed the phase of the fermion determinant such
that $\det\,M_t$ is real. 
However, we still have a sign ambiguity.
Suppose we want to fix the sign but also insist on a smooth gauge
field dependence of $\det\,M_t$. This implies that the sign at $t=0$
fixes the sign along the whole path $\Gamma$. In terms of the angle
$\theta_j$ our numerical result runs as follows: Only the angle
$\theta_0$ crosses zero and changes sign. A glance at \pref{def_detM}
immediately tells us that the fermion determinant changes sign along
the closed path $\Gamma$ and is not a single valued function.


\section*{Conclusions}

Let us give a brief summary of our result.
We have shown that there are closed loops in configuration space
on which the phase $\MT = - 1$.
In a finite lattice it would be possible to define a current
$j_{\mu}(x)$ such that the associated measure produces a Wilson line
${W} = -1$. However, the conflict arises because the proper behaviour in the
classical continuum limit cannot be reproduced:
Going to large physical lattices (small lattice spacing) and insisting
in ${W}=-1$, implies giving up the locality or the gauge invariance
of the theory. Our argument is completely analytical, in particular the
behaviour close to the continuum limit is under control.

We have also checked numerically the original argument given by Witten. 
We find that along a path connecting two gauge fields that differ 
by a topologically non-trivial gauge transformation, the lowest 
eigenvalue of Neuberger's operator crosses zero.
Subsequently there exist closed loops in configuration
space where the fermion determinant changes sign.

On the lattice all gauge transformations can be deformed to the identity
mapping. As we have seen it does not mean that global anomalies are
not present in the lattice formulation, but they rather arise in a
different way. 

\section*{Acknowledgements}

The authors are grateful to Martin L\"uscher for helpful discussions and advise.

The numerical computations have been performed on the Linux cluster
of the DESY-Hamburg Theory Group and on the RTNN Linux farm in Zaragoza
(Spain). 

\newpage 


\end{document}